\documentclass[conference]{IEEEtran}

\usepackage{cite}
\usepackage{amsmath,amssymb,amsfonts}
\usepackage{algorithmic}
\usepackage{graphicx}
\usepackage{textcomp}
\usepackage[dvipsnames]{xcolor}
\def\BibTeX{{\rm B\kern-.05em{\sc i\kern-.025em b}\kern-.08em
    T\kern-.1667em\lower.7ex\hbox{E}\kern-.125emX}}

\usepackage{minted}
\usepackage{listings}
\usepackage{color}
\usepackage{enumitem,amssymb}
\usepackage{bbm}
\usepackage{dsfont}
\usepackage{mathtools}

\usepackage{booktabs}
\usepackage[group-separator={,}]{siunitx}
\definecolor{dkgreen}{rgb}{0,0.6,0}
\definecolor{gray}{rgb}{0.5,0.5,0.5}
\definecolor{mauve}{rgb}{0.58,0,0.82}

\DeclareMathOperator*{\argmin}{argmin} 

\ifCLASSOPTIONcompsoc
    \usepackage[caption=false, font=normalsize, labelfont=sf, textfont=sf]{subfig}
\else
    \usepackage[caption=false, font=footnotesize]{subfig}
\fi

\begin{document}

\title{Unstructured Hydrodynamics on Spatial Dataflow Architectures: A Joint Code and Data Decomposition Approach}

\author{Piotr Luczynski$^{\dagger,\ddagger}$, Tal Ben-Nun$^*$, Leighton Wilson$^\dagger$, Brian Van Essen$^*$\\
$^*$Lawrence Livermore National Laboratory. Livermore, CA, USA\\
$^\dagger$Cerebras Systems. Sunnyvale, CA, USA}

\newcommand\blfootnote[1]{%
  \begingroup
  \renewcommand\thefootnote{}\footnote{#1}%
  \addtocounter{footnote}{-1}%
  \endgroup
}

\maketitle
\global\csname @topnum\endcsname 0
\global\csname @botnum\endcsname 0

\blfootnote{$^\ddagger$ Work was performed during an internship at Lawrence Livermore National Laboratory.}

\begin{abstract}
Spatial Dataflow Architectures are an emerging hardware pattern in high-performance computing, whose mesh-connected fixed-memory processing elements are tailored for structured grid kernels with two-dimensional neighborhoods.
However, practical multiphysics codes are often computed on unstructured grids, which induce indirect memory accesses and high-dimensional communication patterns, making them infeasible to directly map onto said architectures.
This work takes a principled, model-centric approach to partitioning unstructured problems onto spatial dataflow architectures. Through communication and memory modeling, we propose a joint decomposition that considers both the size of the application's fields and its subroutines. In particular, we automate the analysis process of the original code, define a high-dimensional decomposition that minimizes communication via space-filling curves, and apply memory optimization techniques, crucial in this memory-limited environment. We demonstrate mapping the Livermore Unstructured Lagrangian Explicit Shock Hydrodynamics (LULESH) application to the Cerebras Wafer-Scale Engine, showing that larger, unstructured grid codes can still outperform GPUs.
\end{abstract}

\maketitle

\lstdefinelanguage{csl}{
  basicstyle=\fontencoding{T1}\selectfont\ttfamily\scriptsize,
  keywords={@activate,@unblock,@export,@export_symbol, @get_color, @get_dsd, @map, @range, @get_output_queue, @get_local_task_id, @mov16, @fmacs, @bind_local_task, @set_rectangle, @set_tile_code, @set_color_config},
  keywordstyle=\color{blue},
  keywords=[2]{f32, i16, u16, void, WEST, EAST, RAMP, fabout_dsd},
  keywordstyle=[2]\color{MidnightBlue}\bfseries,
  keywords=[3]{var, const, task, fn, for, layout, comptime},
  keywordstyle=[3]\color{Mulberry}\bfseries,
  keywords=[4]{sys_mod,unblock_cmd_stream},
  keywordstyle=[4]\color{Bittersweet},
  numberstyle=\tiny\color{gray},
  commentstyle=\color{dkgreen},
  stringstyle=\color{mauve},
  sensitive=false,
  comment=[l]{//},
  morestring=[b]',
  morestring=[b]"
}

\renewcommand{\listingscaption}{Listing}

\lstset{
   language=csl,
   extendedchars=true,
   basicstyle=\scriptsize\ttfamily,
   numbers=left,
   escapechar=!,
   showstringspaces=false,
   showspaces=false,
   tabsize=2,
   breaklines=true,
   showtabs=false,
   frame=lines
}

\section{Introduction}
\begin{figure}[t]
\centerline{\includegraphics[width=\columnwidth,clip,trim={0cm 0.55cm 0.88cm 0cm}]{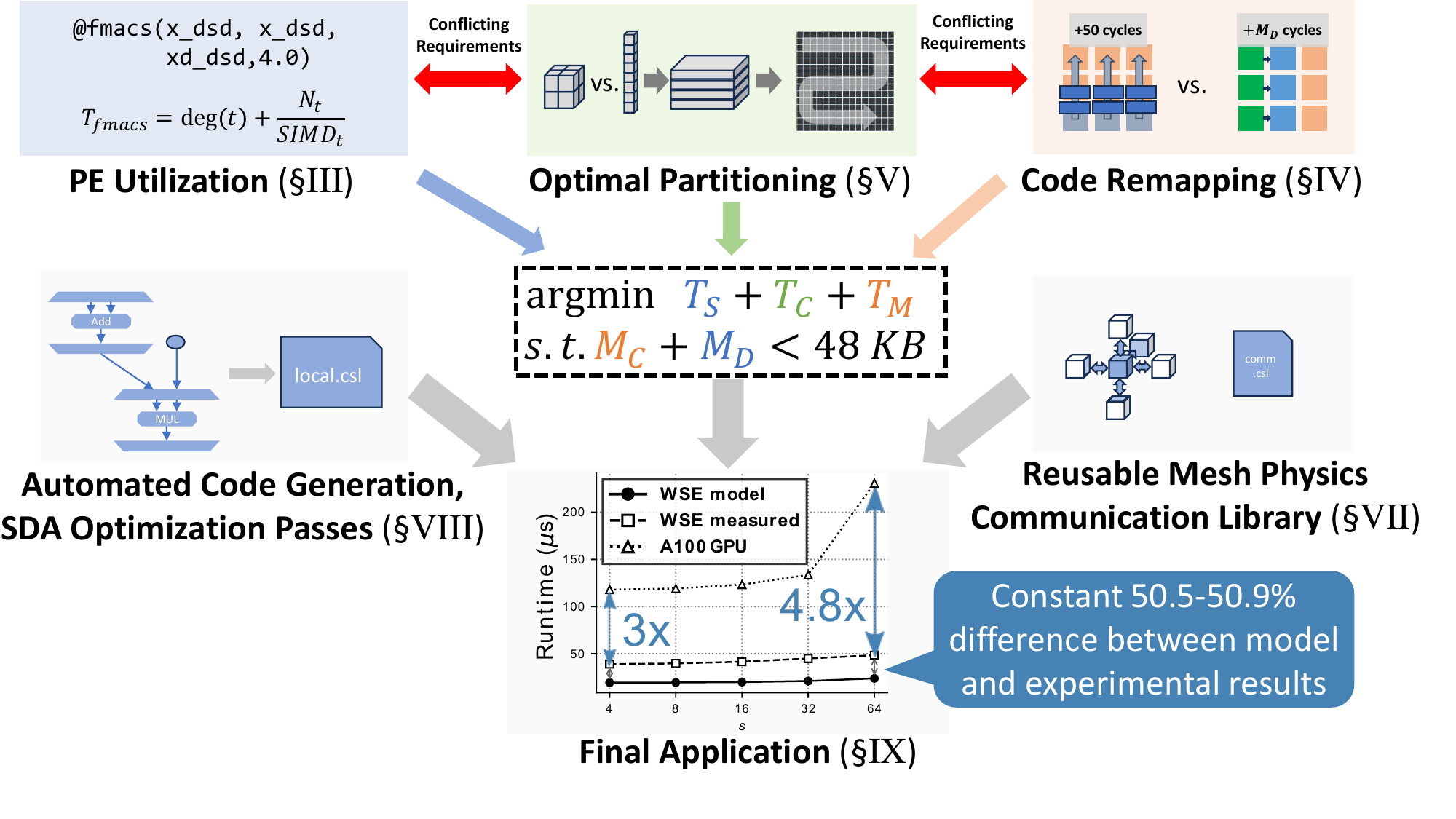}}
\vspace{-0.5em}
\caption{Overview of our approach.}
\vspace{-1em}
\label{fig:overview}
\end{figure}

The growing demand for AI compute has driven the development of a new generation of purpose-built ML accelerators, one of which is the Cerebras Wafer-Scale Engine (WSE)~\cite{benchmarking_wse2}. It is a spatial dataflow architecture that co-locates computation and memory at massive scale, avoiding the off-chip bottlenecks that limit conventional architectures. Its properties have also made it attractive for scientific computing, with demonstrated use across a range of workloads including stencils~\cite{stencil, stencil_2}, seismic processing~\cite{cerebras_seismic}, and molecular dynamics~\cite{cerebras_molecular, cerebras_molecular_2}.

However, these efforts are largely application-specific ports that have not demonstrated generality beyond their target workload. Some rely on manually written assembly~\cite{cerebras_molecular}, which becomes infeasible as application complexity grows. Others target local computations where the domain has a direct mapping onto the 2D PE grid~\cite{stencil}, avoiding the memory pressure and irregular communication patterns that characterize more complex scientific codes. What is missing is a systematic methodology applicable across a broad class of applications.

The WSE poses a set of challenges that are not present on other architectures. First, the per-PE instruction memory is severely limited. For larger scientific codes, a single PE might not be able to store the entire program. Second, unlike CPU or GPU clusters, where network latency is largely uniform, physical distance on the WSE directly determines communication cost. Data partitioning strategies in the 2D space are therefore crucial. Lastly, developers must consider both the domain and code layout while thinking about how it will affect the local computation. This presents itself as a challenging constrained optimization problem.

In this work, we establish a principled approach to mapping scientific applications onto the WSE, using analytical performance and communication models, depicted in Figure~\ref{fig:overview}. The models jointly inform decisions on domain decomposition and code partitioning. We derive these models analytically from NumPy codes leveraging the SDFG IR~\cite{dace}. To support applications, we develop a communication library focused on multidimensionally-decomposed meshes. Finally, we develop a CSL code generator that performs aggressive memory optimizations, generating highly efficient intrinsics that utilize the PE's SIMD capabilities.

We evaluate our approach with the Livermore Unstructured Lagrangian Explicit Shock Hydrodynamics (LULESH) proxy application. Our automatically generated code, combined with our communication library, achieves up to $4.8\times$ speedup over an NVIDIA A100 GPU. Importantly, the measured runtime is within $50\%$ of our model's predictions, validating the model's utility for guiding decomposition decisions.

The paper makes the following contributions:
\begin{itemize}
    \item Performance and communication models that have constant overhead over experimental execution time.
    \item A reusable communication library for mesh-based applications supporting efficient neighborhood exchange across all dimensions of a 3D decomposition.
    \item An automated code generator that performs novel memory optimizations and generates efficient WSE code.
\end{itemize}

\section{Background and Related Work}\label{sec:background}
\subsection{Cerebras Wafer-Scale Engine}

The Cerebras Wafer-Scale Engine (WSE) is a computing architecture first introduced to accelerate machine learning. The novelty of the architecture lies in being able to fit  hundreds of thousands of processing elements (PEs) on a single piece of silicon. At the time, this allowed for fitting large neural networks on a single device, greatly accelerating training time. Recently, we have seen different efforts in using the chip for workloads it was not initially designed for, specifically for more traditional HPC architectures. It has been used to accelerate stencil computations~\cite{cerebras_stencils}, seismic processing~\cite{cerebras_seismic_processing} and molecular dynamics simulations~\cite{cerebras_molecular}.

\begin{figure}[t]
    \centering
    \includegraphics[height=2in,clip,trim={0cm 0.7cm 0cm 0.2cm}]{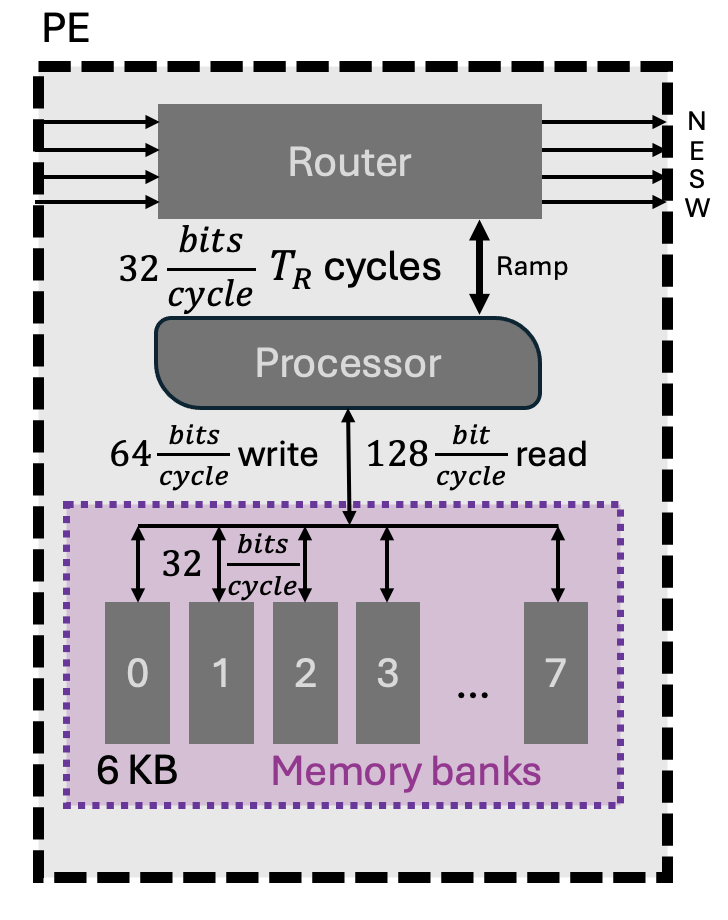}\vspace{-0.5em}
    \caption{Overview of a single PE on a CS-2 chip.}
    \label{fig:pe_overview}
    \vspace{-1em}
\end{figure}

The second-generation WSE, the CS-2, is a 2D mesh of 750,000 user-accessible PEs. Each PE also acts as a router connected to a processor and one level of memory. See Figure~\ref{fig:pe_overview} for an overview of the chip.

The PE has only one level of memory and, unlike traditional CPU architectures, does not contain caches. In each cycle, the processor can read up to 128 bits and write up to 64 bits. However, it can read and write at most 32 bits with a specific bank each cycle. Each of the 8 banks is 6 KB, which amounts to a total of 48 KB. This memory is used to store both data and instructions.

The processor has normal registers to store scalar values. It also has special Data Structure Registers (DSR). Those are used to store information about access patterns, called Data Structure Descriptors (DSD), of an array. DSRs can be used to perform vectorized SIMD operations, depending on whether a \textit{DSD intrinsic operation} exists and provided there are no bank conflicts. Certain computations and inter-PE communication can also run concurrently via a limited number of \textit{microthreads}.

Each PE has a router connected to the processor and \texttt{NORTH}, \texttt{EAST}, \texttt{SOUTH} and \texttt{WEST} PE routers with bidirectional links. Each of those links has a bandwidth of 32 bits per cycle in each direction. Additionally, the link between router and processor (called \texttt{RAMP}) has a latency of $T_R\approx 2$ cycles~\cite{cerebras_reduce}, whereas the links between the routers have only 1 cycle latency. At 1 cycle per hop, the cost to span the entire wafer is 1{,}744 cycles.

The data is sent in 32-bit packets called \emph{wavelets}, each of which is assigned one of 32 colors. A router propagates the wavelet in one or more directions (multicast) based on its color. If two wavelets with the same color arrive at a router at the same cycle this will lead to undefined behavior and needs to be avoided by the programmer.

\subsection{Programming Model}
Running programs on the WSE is supported by the low-level CSL programming language~\cite{cerebras_sdk}. CSL is an extension of the Zig language, which allows the programmer to program both the processor and the router of each PE, exposing various hardware features. CSL code files map to rectangular subsets of PEs, where a layout CSL file maps PEs to code files.

In CSL, there are four types of DSDs:  \texttt{fabin\_dsd}, \texttt{fabout\_dsd}, \texttt{mem1d\_dsd} and \texttt{mem4d\_dsd}. \texttt{fabin\_dsd} and \texttt{fabout\_dsd} describe a sequence of incoming and outgoing wavelets respectively. \texttt{mem1d\_dsd} describe access pattern to a region in memory with some stride. Finally, \texttt{mem4d\_dsd} describes an access pattern to a region in memory that can be defined with up to four variables. Before DSDs can be used by the hardware, they need to be loaded into one of the available DSRs. Below is an example DSD declaration:
\begin{lstlisting}[numbers=none]
const oddElements = @get_dsd(mem1d_dsd, .{
  .tensor_access = |i|{5} -> array[2 * i + 1] });
\end{lstlisting}

Although DSD intrinsics are efficient, they only support elementary operations. For more complex computations, CSL provides a \verb|@map| construct. Given a list of DSDs for each iteration, it will load one value from each one, compute the given function and write a single output to a specified DSR.

\subsection{Performance Modeling of Spatial Dataflow Architectures}
After understanding the hardware, the next step in designing efficient kernels is understanding performance characteristics. Several performance models were proposed for parallel and distributed architectures~\cite{pram_model,bsp_vs_logp_model}, which do not match the asynchronous nature of the WSE. The recent Spatial Computer model~\cite{spatial_computer} was created with spatial dataflow architectures such as the WSE in mind. Unlike other models, it takes the distance of messages into account, which for the CS-2 is crucial.

A modified version of the model has been used to accurately predict the performance of communication collectives for CS-2~\cite{cerebras_reduce}. To predict communication cost, it uses a metric called \emph{dependent distance} ($L$), which is the largest number of hops any wavelet needs to be routed through for the algorithm to execute. It also considers \emph{aggregated distance} ($E$), which is the total number of hops of all wavelets. Additionally, it considers the algorithmic \emph{depth} ($D$), which is the longest sequence of wavelets that depend on each other. Finally, it considers \emph{contention} ($C$) which is the largest number of incoming or outgoing wavelets for any PE. Given $N$ links, a lower bound on the time to compute an algorithm is approximated as:

$$T = \max(C, \frac{E}{N} + L) + 2(T_R + 1)$$

\subsection{LULESH}
\begin{figure}[t]
    \centering
    \begin{minipage}[t]{0.49\columnwidth}
        \centering
        \includegraphics[width=\linewidth]{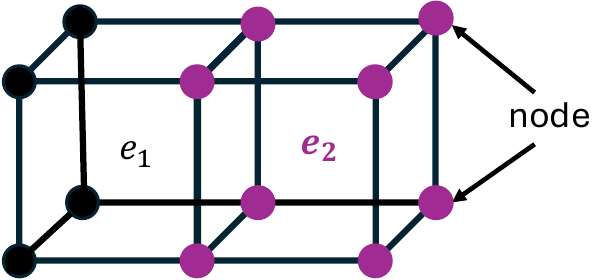}
        \caption{Two elements in LULESH and their surrounding nodes. The purple nodes define element $e_2$.}
        \label{fig:element_and_nodes}
    \end{minipage}
    \hfill
    \begin{minipage}[t]{0.35\columnwidth}
        \centering
        \includegraphics[width=\linewidth]{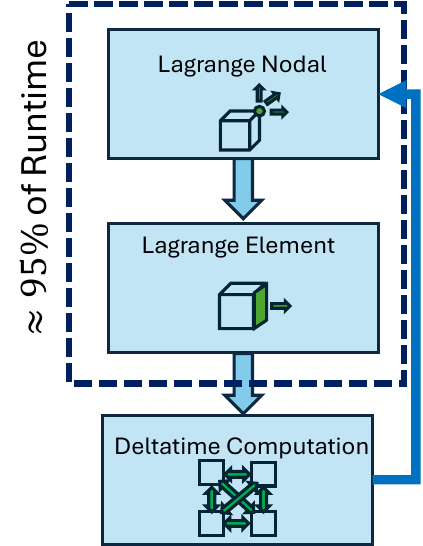}
        \caption{LULESH loop diagram with top-level procedures. The diagrams depict the communication type in each kernel.}
        \label{fig:lulesh_overview}
    \end{minipage}
    \vspace{-1em}
\end{figure}

LULESH~\cite{lulesh} is a proxy application developed by Lawrence Livermore National Laboratory as a representative for the computational patterns of a typical hydrodynamics code. It is a Lagrangian simulation operating on a hexahedral mesh of elements (Figure~\ref{fig:element_and_nodes}). Each element cell contains properties such as pressure or internal energy. The shape of an element is defined by its 8 nodes, which have properties such as position or force.

In a distributed environment, each process will store a subset of the elements in the whole domain including the nodes that define them~\cite{lulesh, lulesh_model_comparison}. In those implementations, if we have $e_1$ be in $p_1$ and $e_2$ in $p_2$, the bordering nodes would be replicated at each process.

\subsubsection{Kernel Characterization}
Over the runtime of the application we execute different types of kernels. In Figure~\ref{fig:lulesh_overview} we provide an overview of the whole application. We also overlay the percentage of the total runtime when the application is run with OpenMP. The green shapes represent communication points in a distributed environment.

The most common kernels are node-wise or element-wise kernels, iterating over each element or node, respectively, and their properties in parallel. 
Another type of kernel are those which look at the neighborhood of an element. Specifically, they consider the 8 nodes surrounding an element. They use both the surrounding nodes and the element to compute a property that belongs either to an element or a node. 

Notice that if we update a property of a node, it will be different based on the element-neighborhood chosen. Because some nodes might be replicated at multiple processors, this requires us to communicate the updated value. 
If we are exchanging node properties, we will need to communicate with $26$ other processors. 

There are also kernels that require properties of neighboring elements. Those elements might not be stored at the given processor. This inserts another communication step marked with a square. In this case we only need to communicate with processors bordering on the faces, which there are 6.

Finally, at the end of each time-step loop we need to perform an \texttt{allreduce} to compute a new value of $\Delta t$.

\subsection{Stateful Dataflow Multigraphs}

We would like to perform our analysis and optimizations in a verifiable and holistic manner. Since most of the analysis revolves around dataflow, we choose the stateful dataflow multigraph (SDFG)~\cite{dace} graph-based intermediate representation (IR). 

\begin{figure}[t]
    \centering
    \includegraphics[width=0.6\linewidth]{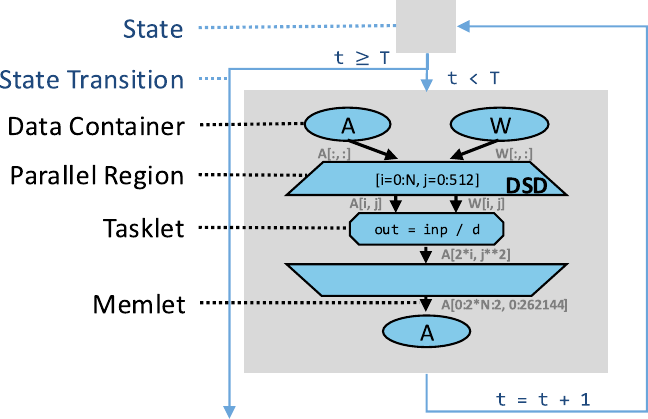}
    \caption{Schematic overview of a Stateful Dataflow Multigraph (SDFG).}
    \label{fig:sdfg}
\end{figure}

The SDFG (see Figure~\ref{fig:sdfg}) is a directed graph of acyclic dataflow multigraphs, where the former represents a control flow state machine and the latter a dataflow graph. SDFGs keep global knowledge of data containers (arrays, queues, views) and code is executed in fine-grained \textit{tasklet} nodes. Tasklets are not allowed to access memory not prescribed by edges in the dataflow graph, called \textit{memlets}, which represent constrained, bounded-volume units of data movement. Memlets are propagated through parallel regions (\textit{Maps}) and loop scopes, so the entire accessed region is known. Additionally, SDFGs by design maintain an $\mathcal{O}(1)$ data access provenance, which allows compiler passes to reason about which data is being accessed, including through reshapes and references.

The SDFG IR powers libraries such as DaCe and integrates with compiler infrastructure such as MLIR~\cite{mlir}. DaCe offers a set of passes that complement control-centric passes found in compilers such as LLVM~\cite{llvm}, including array elimination and memory reuse, which are crucial for resource-constrained environments. NumPy codes can be directly ingested into SDFGs, which further motivates our use of the representation with the PyLULESH~\cite{pylulesh} port of LULESH.

In this work, we identify that the SDFG provides a powerful representation that is well-aligned with the WSE hardware architecture. For example, parallel regions (under the right, analyzable conditions) can be scheduled to DSD intrinsics, \texttt{@map}s, or loops. With the optimizations and code generator proposed in Section~\ref{sec:compiler}, SDFGs enable new classes of scientific applications to be efficiently mapped onto the WSE.

\section{Local Compute and Memory}
\label{sec:compute}
Our goal is to maximize WSE PE utilization by leveraging microthreads and SIMD operations via DSDs. On the WSE, microthreads allow for concurrent but not parallel execution, which allows us to model all operations directly with a work-depth model. The model will have two parts, data staging and computation, both of which consume cycles. Unlike CPUs and GPUs, which rely on DRAM, operations on SRAM consume a similar number of cycles for both compute and data access.

Because we will later use the model to guide the distributed domain decomposition in combination with a communication model, we need to develop a model that is parameterized with respect to the local domain characteristics.
The goal is not to develop an exact, cycle-accurate model, but rather a close approximation to help guide implementation choices. We will conclude with a model of the form:
$$T_{local}(e_x, e_y, e_z) \text{ and } I_{local}(e_x, e_y, e_z)$$
for runtime and number of instructions respectively.

\subsection{Gauging the Effectiveness of DSDs via Microbenchmarking}
Before developing the model, we first need to understand the performance characteristics of different instructions. Often there are multiple ways to implement the same operation. A loop can also be implemented as a CSL \texttt{@map} or a DSD intrinsic under certain conditions. Appropriate modeling of the ratios between the operations' cycle counts helps guide implementation and algebraic design decisions (e.g., fusion vs. sequence of intrinsics). We examine two common patterns in LULESH and measure the performance of different implementations.

We will consider two functions, where the first one is a simple fused multiply-add and the second one is an indirect access. The latter is more difficult to implement on the hardware as we cannot use a DSD intrinsic.

\subsubsection*{Fused Multiply-Add}
The FMA operation is given as
$$x := x + x_d \cdot \Delta t,$$

where $x$, $x_d$ are node properties and $\Delta t$ is a scalar. In the following microbenchmark, we assume $x$ and $x_d$ are vectors of length $6^3$. Below are several CSL implementation approaches:
\begin{lstlisting}[numbers=none]
// Loop
for (@range(u16, 216)) { d = a * b + c; }
// Map
fn mac(a: f32, b: f32, c: f32) f32 { return a * b + c; }
@map(mac, yd_dsd, deltatime, y_dsd, y_dsd);
// DSD Intrinsic
@fmacs(x_dsd, x_dsd, xd_dsd, deltatime);
\end{lstlisting}

We evaluate the runtime of these implementations, reporting the latency and code size in Table~\ref{tab:fma}.

\begin{table}[t]
\caption{Runtime and Code Size of Fused Multiply-Add}
\begin{center}
\begin{tabular}{lrr}
\toprule
\textbf{Implementation} & \textbf{Runtime [cycles]} & \textbf{Code Size [bytes]} \\
\midrule
For Loop       & 7139                 & 128 \\
Map            & 1525                 & 72  \\
DSD     & \textbf{291}         & 48  \\
\bottomrule
\end{tabular}
\label{tab:fma}
\end{center}
\end{table}

The performance differences are substantial, with the DSD implementation almost 30$\times$ faster than a for-loop. Inspecting the for-loop generated assembly, we can see that much of the code inside the loop is spent on loading data, with multiple instructions used to access a single element. This overhead is reduced in a map, where each data load takes 1 cycle because it is defined by a DSD. Finally, using a DSD intrinsic, we are able to execute almost one fused multiply-add per cycle.

\subsubsection*{Indirect Access}
A common pattern in LULESH is gathering the nodes neighboring an element. In arbitrary meshes, such operators do not follow a regular pattern. In CSL:

\begin{lstlisting}[numbers=none]
for (@range(u16, 8)) |i| { xlocal[i] = x[nodelist[i]]; }
\end{lstlisting}

The operation can be implemented using a map as:

\begin{lstlisting}[numbers=none]
fn collect_x(idx: u32) f32 { return x[idx]; }
@map(collect_x, nodelist_dsd, xlocal_dsd);
\end{lstlisting}

Unless the values of \texttt{nodelist} are known in advance, this pattern cannot be implemented with DSD intrinsics. We benchmark the implementation and report the results in Table~\ref{tab:indirect}. As in the previous experiment, the map-based approach performs better in both runtime and memory use.

\begin{table}[t]
\caption{Indirect Access: runtime and code size per implementation.}
\begin{center}
\begin{tabular}{lcc}
\toprule
\textbf{Implementation} & \textbf{Runtime [cycles]} & \textbf{Code Size [bytes]} \\
\midrule
For Loop & 224 & 64 \\
Map      & \textbf{81}  & 60 \\
\bottomrule
\end{tabular}
\label{tab:indirect}
\end{center}
\end{table}

We conclude that, in order to utilize the WSE to its fullest extent, we should use DSD operations where they are supported. Second, the number of instructions and performance are highly correlated due to hardware characteristics.

\subsection{Estimating Procedure Performance and Memory}
Now that we have seen the performance characteristics of different CSL instructions, we would like to estimate the performance of any program represented as a stateful dataflow multigraph (SDFG). Consider an SDFG \textit{map} containing a single \textit{tasklet} $t$. Let $N_t$ be the number of times this tasklet is executed, i.e., the size of the map. Let $\text{SIMD}_t$ be the supported SIMD width for that tasklet (computed based on its constituent operators). Let $deg(t)$ be the degree of the tasklet node, i.e., the sum of its incoming and outgoing dataflow edges (\textit{memlets}). Let $v_t \in \{0, 1\}$ denote whether the tasklet corresponds to a supported DSD intrinsic operation. Below, we derive a cost $T(t)$ and instruction count $I(t)$ for each case.

\subsubsection{$v_t = 1$} If the operation in the tasklet is supported by a DSD operation, we can vectorize it. In that case, execution takes $\left\lceil\frac{N_t}{\text{SIMD}_t}\right\rceil$ cycles. However, a setup cost is also incurred: WSE setup requires setting the DSR addresses for the inputs and outputs, which takes at least $deg(t)$ cycles. Therefore:

$$T(t) = deg(t) + \left\lceil\frac{N_t}{\text{SIMD}_t}\right\rceil.$$

The number of instructions would be at least:
$$I(t) = 1 + deg(t)$$

\subsubsection{$v_t = 0$} If the operation cannot be vectorized, we execute the tasklet $N_t$ times, with each execution taking \emph{at least} one cycle, giving $N_t$ cycles. We also need to account for loads and stores to the appropriate registers. We assume each distinct address requires a separate load or store, and let $D_t$ denote the total data volume accessed by the tasklet, obtained by summing the volume of each memlet. Each data access costs $1$ cycle on this architecture. This results in a runtime of:

$$T(t) = N_t + D_t$$

In terms of code size, we need $1$ instruction to load each input and output of a tasklet and $1$ instruction to execute it. If $N_t > 1$, we also need at least one instruction to jump back to the start of the loop. Note that CSL supports loop implementations that do not check a stopping condition on every iteration. We account for the jump in the instruction count but not in the performance model, since the loop could in principle be fully unrolled. This gives:

$$I(t) = \mathbf{1}_{N_t > 1} + 1 + deg(t)$$

Note that there are several factors we do not model: loop unrolling affects code size but improves performance, and some loads may be unnecessary if values are already in the correct registers.

\subsection{Automated Dataflow Graph Analysis}
Using the model above, we design a fully automated pass that estimates the runtime of a stateful dataflow multigraph. The pass automatically analyzes the number of SDFG tasklets, predicted performance, and code size. It is general and can be applied to various programs, with the analysis parameterized with respect to the program input sizes. The total performance and code size of an SDFG are obtained as:

\begin{align*}
T_{local}(e_x, e_y, e_z) &= \sum_{t \in \text{SDFG}_{e_x,e_y,e_z}} T(t) \\
I_{local}(e_x, e_y, e_z) &= \sum_{t \in \text{SDFG}_{e_x,e_y,e_z}} I(t)
\end{align*}

As WSE processing elements execute instructions serially, the model assumption is that from a Work-Depth perspective, the depth is equal to the work (i.e., the intra-PE average parallelism is 1).

We now use this approach on LULESH to create a parameterized model of its two top-level computational functions. Let $n_f = 2 n_x n_y + 2 n_x n_z + 2 n_y n_z$ be the number of face nodes on the boundaries of the mesh. If we consider that SIMD operations are supported for lengths of at least 4, the programs can be expressed as:
$$T_{nodal} = \begin{cases} 7603 \cdot e + 96 \cdot n + n_f + 933 & e < 4 \\ 4297 \cdot \lceil e/2 \rceil + 21 \cdot n + 4545 & e \geq 4 \end{cases}$$

$$T_{element} = \begin{cases} 6267 \cdot e + 84 & e < 4 \\ 3765 \cdot \lceil e/2 \rceil + 3879 & e \geq 4 \end{cases}$$

The computation is mostly linearly dependent on the number of elements, with boundary conditions and node-based computations contributing a minority of the cost. In the vectorized version we have a larger constant factor which corresponds to the DSD setup costs. In the non-vectorized version, we have a larger multiplier with respect to the program parameters, which is due to both explicit data movement and lack of vectorization. In this case, using DSDs for larger domain sizes can help us significantly improve performance.

Our approach also allows us to estimate $I_{local}$, which is large: the local code alone is estimated to require upwards of 24~KB. This must be balanced against data and communication code to fit within the 48~KB total memory budget.

\section{Code-Aware Memory Model}
\label{sec:memory}

\subsection{Memory Requirements}
To decide how to decompose the program between the PEs, we need to consider how large a part of the domain can be stored in each PE. We need to know whether we are able to allocate more than a single element per PE. Let $M_D$ and $M_C$ be the memory required to store all the needed data and all the instructions, respectively. Let us first see how $M_D$ changes based on the local number of elements.

Let $e$ be the number of elements and $n$ the number of nodes per PE. The total memory requirement to store the domain is:

$$M_D = 4\cdot (28 \cdot e + 13 \cdot n)$$

This allows us to store at most $e = 216$ elements for which $M_D =$ 42~KB. Note that $n \geq (\sqrt[3]{e} + 1)^3$.

This, however, does not account for the code size, which can be at most $HW_C=$32~KB per PE. As discussed in the previous section, according to our memory model we would need more than 24~KB just for local instructions. Additionally, we measure the communication code size to be 9~KB. This means that $M_C > HW_C$ and the whole program does not fit on a single PE together with the domain requirements. This problem is exacerbated as application size grows.

\subsection{Mapping Approaches}
\begin{figure}[t]
\centerline{\includegraphics[width=\columnwidth]{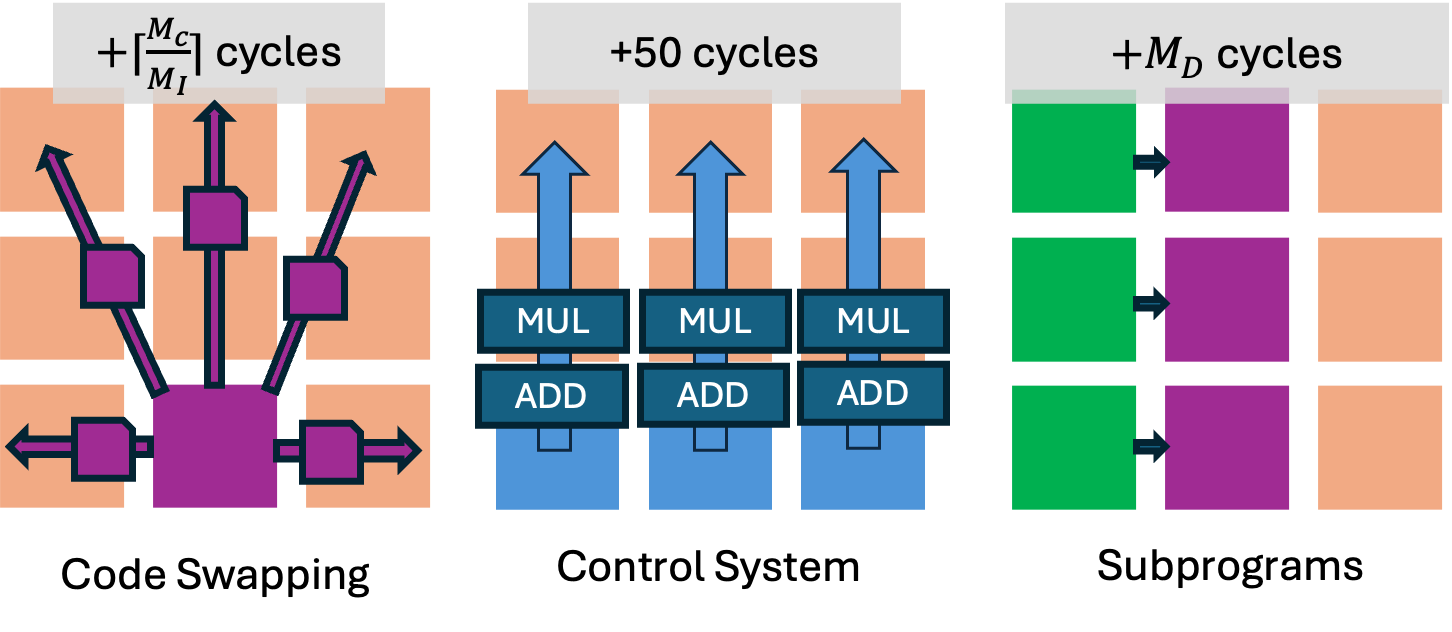}}
\caption{Different approaches for mapping code with their respective costs.}
\label{fig:memory_mapping}
\end{figure}
The inability to fit code on a single PE is not LULESH-specific; it is expected to be a problem for any large scientific application. There are a couple of possible solutions if we run out of instruction space. Specifically, the alternatives are code swapping, using a control system~\cite{control_system}, or mapping subprograms on multiple PEs. We model the costs in a general setting and discuss the trade-offs of each of those approaches. 
See Fig.~\ref{fig:memory_mapping} for an overview. 

\subsubsection*{Code swapping} an approach where code on the PE is overwritten/reconfigured at runtime. It could be considered a lighter version of context switching~\cite{context_switching, context_switching_2} since the data fields remain in the PE. The number of times we need to swap the code would ideally be $\left\lceil\frac{M_C}{HW_C}\right\rceil$. The total overhead this would incur is at least 
$$T_{swap} \ge \frac{M_C}{4} \text{ cycles.}$$
This is because each wavelet can carry $4$ bytes, and we do not consider chip-wide synchronization costs (which can be modeled as a barrier or allreduce).
\subsubsection*{Control system} a control system abstracts basic operations (such as Addition or Multiplication) to specific tasks. Then, a controller sends commands to activate these tasks with specific parameters across the chip. This allows us to significantly reduce code space by keeping only general kernels on each PE and outsourcing the control overhead. However, it comes with a trade-off where each such command might incur a 50-60 cycle latency needed to load the 
arguments~\cite{control_system}. Let us model a single overhead $O$ for each command (this is not fully representative since such overhead would depend on the command); the total overhead can be estimated as:
$$T_{control} = \sum_{t \in \text{SDFG}} O  \text{ cycles.}$$ 
To mitigate the costs, common subprograms can be fused into coarser-grained tasks, which in turn costs additional $M_C$.

\subsubsection*{Subprogram PEs} this approach separates the program into disjoint parts and assigns them to different PEs. The PEs can be mapped to a local topology, such as a braided ring~\cite{connection_machine}, and executed sequentially. Let $S_i$ be the $i$-th subprogram PE in the ring. $S_0$ starts, and upon completion of the computation propagates the necessary data to subsequent subprogram PEs. As opposed to the other approaches, one benefit of subprograms is decentralized, asynchronous execution.

Because the data are mapped onto a ring, the distance between two consecutive PEs (including cycling back to $S_0$) is at most 2. If we are sending a vector of length $B$ from $S_i$ to $S_{i + 1}$, we obtain a distance of $2$, message size of $2 \cdot B$ and link contention of $B$. This yields a runtime of approximately $2 + B + 2T_R + 1$ cycles.
We will need a total of $k_p = \left\lceil\frac{M_C}{HW_C}\right\rceil$ PEs and $B\le M_D$, leading to an estimated runtime of:
$$T_{subprogram} \le k_p \cdot (2 + M_D + 2T_R + 1)\text{ cycles.}$$
This is an upper bound, since we only need to be sending an intersection of the data required by the source and sink PEs.

\subsection{Comparison}
\begin{figure}[t]
    \centering
    \includegraphics[width=\columnwidth]{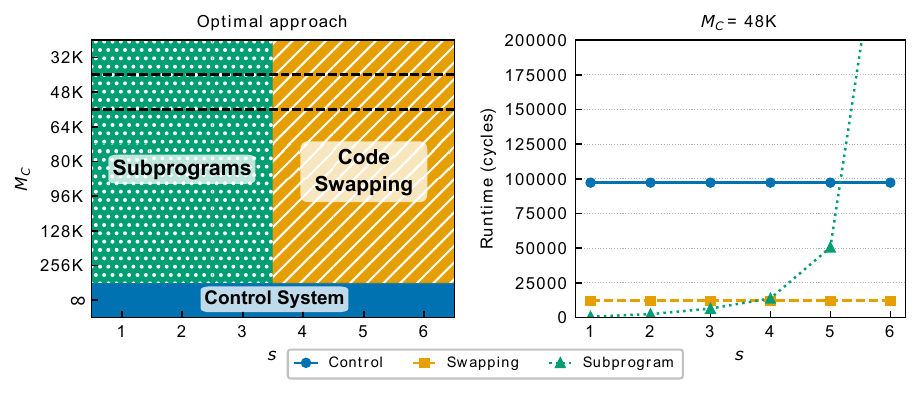}
    \caption{Optimal memory mapping approach by $M_C$ and local domain size $s=\sqrt[3]{e}$ (left), and predicted runtime overhead at $M_C=48$~KB (right).}
    \label{fig:memory_combined}
\end{figure}
Comparing these approaches in our LULESH-specific environment, let us assume that each PE comprises 48~KB and that $M_D = 4 \cdot (28\cdot e + 13\cdot n)$. Since each PE has 48~KB total, the memory available for instructions is $M_I = 32\text{~KB} - \max(0, M_D - 16\text{~KB})$: when the domain data fits within 16~KB the full 32~KB is available for code, but for larger domains the data spills into the instruction budget. From LULESH's dataflow graph we also know that the total number of tasklets is 2{,}430.

We plot a heatmap of the optimal approach for LULESH over various combinations of $M_C$ and $s$, the local domain size, in Fig.~\ref{fig:memory_combined}. The figure shows that for small local domain sizes, using subprogram PEs is optimal. Because the control system approach is constant for a given program, it becomes optimal for very large code sizes. We also show the predicted overhead each approach incurs for $M_C = 48$~KB, which is what we expect for LULESH. In this case we will use either subprograms or code swapping depending on the local domain size we choose.

\section{Communication Model}
\label{sec:comm}
Mesh-based scientific applications such as LULESH are characterized by local communication patterns: each PE exchanges data only with its immediate neighbors in the physical domain. To predict and optimize performance, we need to develop a model of all communication across PEs as a function of the decomposition. We use the Spatial Computer model~\cite{spatial_computer} and show how to instantiate it for LULESH's two communication patterns: allreduce and exchange with face neighbors across all three dimensions ($x$, $y$, $z$, see Table~\ref{tab:comm_symbols} for all symbols). We conclude with a model $T_{Comm}(s, e_x, e_y, e_z, k_p, p_x, p_y)$.

\subsection{Allreduce}
In LULESH, we perform scalar allreduce twice per iteration. The collective has been widely discussed and implemented for various applications~\cite{stencil, stencil_2}. Since this is a known operation, we will use the reported lower bound for our analysis:

$$T_{Allreduce} = P + 2T_R\text{ cycles.}$$

Notice that the runtime is only dependent on the dimensions on which we map the program. Therefore, to optimize for it we need to minimize the maximum distance between any PEs in the program. Additionally, the runtime does not depend on how we map the domain.

\begin{table}[t]
\caption{Communication Model Symbol Definitions}
\begin{center}
\begin{tabular}{l p{0.75\columnwidth}}
\toprule
\textbf{Symbol} & \textbf{Definition} \\
\midrule
$s$ & Side length of the simulated domain; total elements: $s^3$ \\
$e_x, e_y, e_z$ & Number of local elements in $x$, $y$, $z$ directions per PE \\
$n_x, n_y, n_z$ & Number of local nodes in $x$, $y$, $z$ directions per PE;\newline $n_x = e_x + 1$ \\
$k_p$ & Number of subprogram PEs per domain tile \\
$p_x, p_y$ & Wafer dimensions; must satisfy $e_x e_y e_z \cdot p_x p_y / k_p \geq s^3$ \\
$P$ & Total PEs: $p_x \cdot p_y$ \\
$P_k$ & PEs per tile: $p_x p_y / k_p$ \\
$e$ & Elements per PE: $e_x \cdot e_y \cdot e_z$ \\
$W, H$ & CS-2 maximum width and height \\
\bottomrule
\end{tabular}
\end{center}
\label{tab:comm_symbols}
\end{table}

\subsection{Face Communication}
Communication with face neighbors is highly affected by domain decomposition. 
We need to find a 2D mapping that allows us to efficiently communicate with all neighbors in 3 dimensions. We can choose to tile the Z dimension. This will enable the X and Y neighbors in the logical space to also be neighbors in the physical space. However, this does not yet yield any guarantees on the physical distance between the Z neighbors.

\begin{figure}[t]
\centerline{\includegraphics[width=\columnwidth]{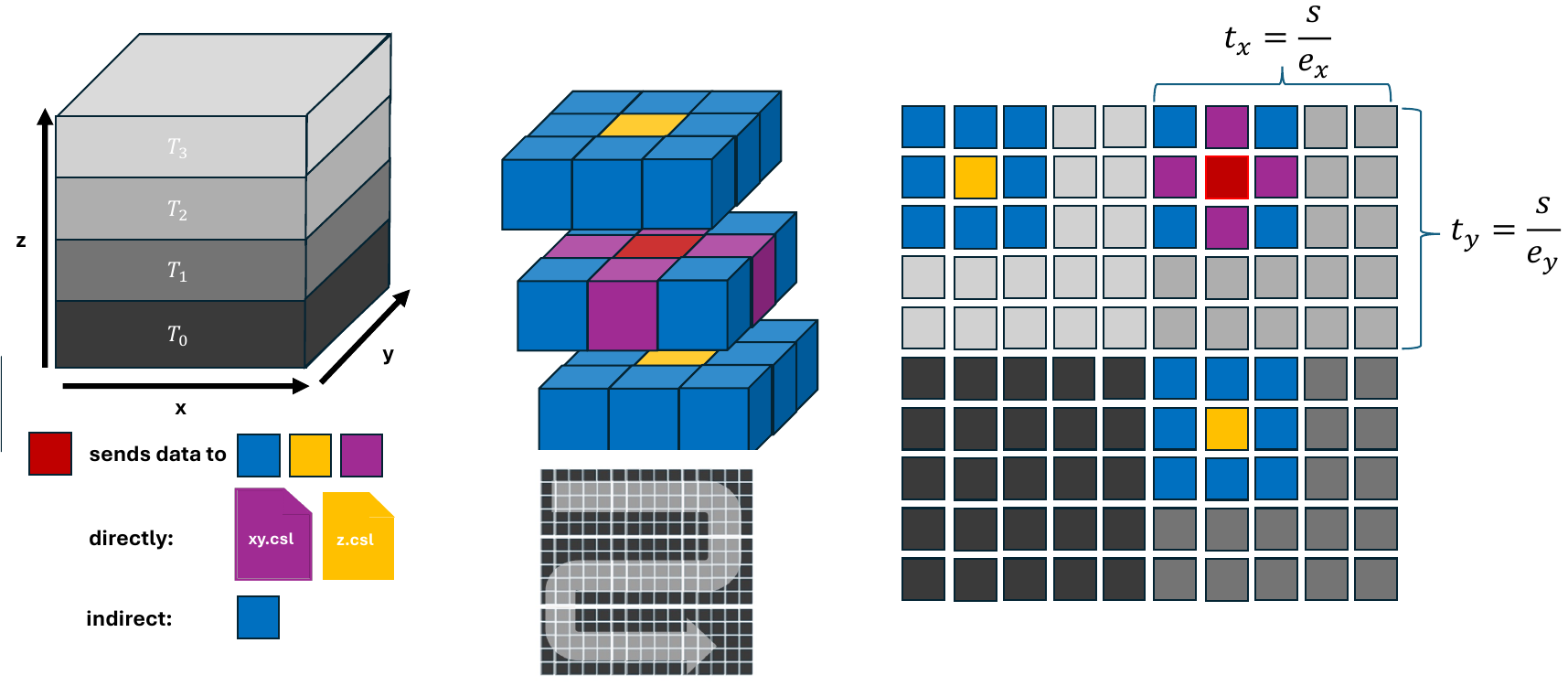}}
\caption{Overview of the 3D domain decomposition approach.}
\label{fig:domain_decomposition}
\end{figure}

To map the Z tiles, we can use space-filling curves~\cite{space_filling_1}, which have been previously explored in HPC applications~\cite{space_filling_hpc_1, space_filling_hpc_2}. We decide to use a snake-shaped space-filling curve, illustrated in Fig.~\ref{fig:domain_decomposition} (bottom center), which will allow Z tiles to neighbor each other, hence bounding the physical distance between the  logical neighbors to be the size of the tile. In our case, it does not form a Hamiltonian cycle, but we could map it in that way if the domain were periodic for some other application.

Let us define $T_x(n), T_y(n), T_z(n)$ as the cost of exchanging $n$ values with a neighbor on one of the planes, e.g., all PEs sending to their next neighbor on the $x$ plane. Communication rounds are sequential, meaning that the cost of exchanging $n$ elements with all neighbors is $2\cdot T_x(n) + 2\cdot T_y(n) + 2\cdot T_z(n)$. In our analysis we assume all PEs are equal and ignore those on the boundary that do not send any messages since either way we are limited by the critical path.

In the subprogram code decomposition case, the dependent distance $L_x$ is equivalent to $k_p$, since the subprograms are neighboring. The depth $D_x$ is 1. Since we send $n$ messages, the contention $C_x$ is $n$. Each of the $P$ PEs sends $n$ messages at distance $k_p$ giving us the aggregate distance $E_x$ as $n \cdot k_p \cdot P$. We can approximate the number of links as $k_p \cdot P$, yielding:

$$T_x = n + k_p + 2\cdot T_R + 1$$

Because $k_p$ is small, it is a negligible constant factor. In the $y$ direction, PEs are physically adjacent ($k_p = 1$), so $T_y = n + 1 + 2\cdot T_R + 1$. In both cases, the cost is dominated by the amount of data sent.

The more complex communication pattern is exchanging data between tiles. The adjacent Z tile might be neighboring horizontally or vertically, hence the distance $L_z$ is $\max(t_x \cdot k_p, t_y)$. Since we are sending $n$ values, link contention $C_z$ is again $n$. For computing the aggregated distance, we will assume the worst case where each PE is sending at the maximum distance $L_z$. This gives us $E_z = P\cdot \max(t_x \cdot k_p, t_y) \cdot n$ as an upper bound. Finally we can approximate the number of links $N_z$ as $k_p \cdot P_k$, per PE we have one link from it to somewhere else at most. This gives us:

$$T_z = \frac{n \cdot \max(t_x \cdot k_p, t_y)}{k_p} + \max(t_x \cdot k_p, t_y) + (2T_R + 1)$$

\subsection{Full Model}
Putting it all together, we execute the communication twice, once exchanging node properties, and another time element properties. In each of the cases, we are sending different amounts of data, which yields a total communication cost of:
\begin{align*}
T^{elem}_{comm} &= 2\cdot T_x(e_ye_z) + 2\cdot T_y(e_xe_z) + 2\cdot T_z(e_xe_y) \\
T^{nodal}_{comm}   &= 2\cdot T_x(n_yn_z) + 2\cdot T_y(n_xn_z) + 2\cdot T_z(n_xn_y)
\end{align*}

Interestingly, for fixed $s, e_x, e_y, e_z$ to minimize the communication cost we need to minimize:
\begin{align*}
\argmin_{p_x, p_y} \quad & 2(p_x k_p + p_y) \\
\text{s.t.} \quad & p_x k_p < W, \quad p_y < H, \\
                  & \left\lfloor \frac{p_x}{t_x} \right\rfloor \left\lfloor \frac{p_y}{t_y} \right\rfloor \geq t_z
\end{align*}

The total communication cost per iteration is then:
$$T_{Comm} = T^{elem}_{comm} + T^{nodal}_{comm} + 2\cdot T_{Allreduce}$$
Given the optimal $p_x, p_y$ from the minimization above, this expression can be evaluated for any decomposition $(s, e_x, e_y, e_z, k_p)$. Combined with the compute and memory models, it forms the complete performance model used to select the optimal decomposition in the next section.
\section{Joint Data and Code Decomposition}
\label{sec:joint_decomposition}
In previous sections we created local compute, communication and memory models, which we now combine. We use them to find the optimal code and data decomposition for given domain size. Let us define $M_P$ to be the maximum memory across all PE classes for both code and domain. We are faced with a constrained optimization problem:
\begin{align*}
\min \quad & T_{Seq}+T_{Comm}+T_{Map}\\
\text{s.t.} \quad & M_C+M_D < 48\text{~KB},
\end{align*}
where $T_{Seq}$ is the sequential computation runtime, $T_{Comm}$ is the lower bound of the communication time, and $T_{Map}$ is the cycle overhead required to map overflowing code to the PEs.

Specifically, for a given domain size $s$ and estimated instruction memory $M_C$, we optimize over $e_x$, $e_y$, $e_z$ and the code mapping approach. The wafer dimensions, which govern overall chip utilization (and thus effective peak performance), follow from that. Since $e_x, e_y, e_z \leq 6$ there is a small number of combinations that we can enumerate for a given domain size to find the optimal mapping.

\begin{table}[t]
\caption{Optimal mapping parameters and predicted runtime}
\begin{center}
\begin{tabular}{rrrrrlr}
\toprule
\textbf{$s$} & \textbf{$e_x$} & \textbf{$e_y$} & \textbf{$e_z$} & \textbf{$k_p$} & \textbf{Approach} & \textbf{Runtime} \\
\midrule
  \textbf{4} & 1 & 1 & 1 & 2 & Subprograms & 16{,}105 \\
  \textbf{8} & 1 & 1 & 1 & 2 & Subprograms & 16{,}257 \\
 \textbf{16} & 1 & 1 & 1 & 2 & Subprograms & 16{,}625 \\
 \textbf{32} & 1 & 1 & 1 & 2 & Subprograms & 17{,}585 \\
 \textbf{64} & 1 & 1 & 1 & 2 & Subprograms & 19{,}953 \\
\midrule
128 & 3 & 1 & 1 & 1 & Swapping    & 62{,}704 \\
256 & 4 & 2 & 3 & 1 & Swapping    & 290{,}229 \\
\bottomrule
\end{tabular}
\end{center}
\label{tab:optimal_mapping}
\vspace{-1em}
\end{table}

See Table~\ref{tab:optimal_mapping} for results. As expected, using subprogram PEs is optimal for most domain sizes, whereas for larger domain sizes code swapping is better. Additionally, as long as it is feasible, keeping the local domain as small as possible, i.e., $e_x = e_y = e_z = 1$, is desirable. This is because most of the runtime is dominated by local computations. Therefore, for our implementation, we will focus on this scenario and implement it for $s \leq 64$.

We will now shift our focus to the implementation. In this work, we provide two general-purpose code components:  Section~\ref{sec:commlib} details the implementation of an optimized communication library, and Section~\ref{sec:compiler} discusses our code generation framework based on the SDFG IR, including WSE-specific compiler passes and a CSL code generator.
\section{Communication Library}\label{sec:commlib}
We introduce a reusable communication library to aid the implementation of 3D mesh physics simulations. The library supports exchange of data between the face neighbors across all three dimensions when the Z dimension is tiled with a space-filling curve. It is currently tailored to LULESH and its communication patterns, but could be easily modified to support other codes.

\subsection{Communication Implementation}
To properly implement communication, we handle both the neighbors on the same tile and across tiles.

\subsubsection*{XY Communication (intra-tile)}
XY communication is straightforward, since each PE communicates with neighbors that are at most $k_p$ hops away. Communication is scheduled sequentially: all PEs send and receive for a single direction before advancing to the next. A PE proceeds to the next direction only once it has both sent all outgoing data and confirmed the data has entered the network. Communication on the chip-wide network is by design one-sided -- a PE cannot wait for the remote PE to have received it. To overlap sending and receiving, we launch two microthreads that alternate: one handles incoming wavelets while the other sends outgoing ones. Sequential scheduling also allows careful virtual channel (i.e., color) minimization, which is important as colors are a scarce resource.

\subsubsection*{Z Communication (cross-tile)}
\begin{figure}[t]
    \centering
    \begin{minipage}[b]{.48\columnwidth}
    \includegraphics[width=\columnwidth]{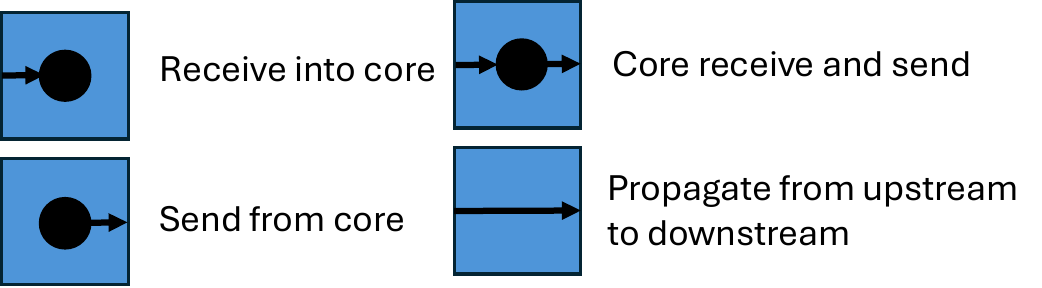}
    \vspace{2em}
    \end{minipage}
    \hfill
    \begin{minipage}[b]{.48\columnwidth}
    \includegraphics[width=\columnwidth]{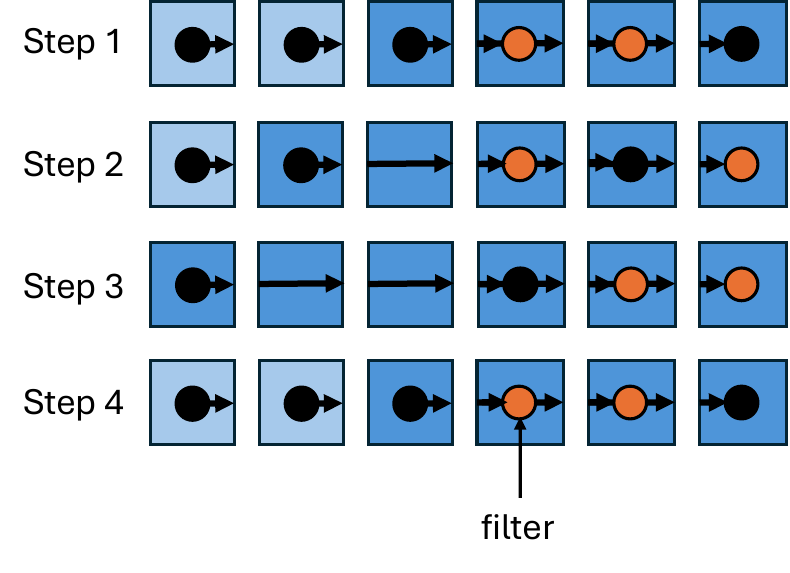}
    \end{minipage}
    \vspace{-1em}
    \caption{Cross-tile communication diagram. A control wavelet at the end of each step advances the filter counter so the next PE becomes the active receiver.}
    \label{fig:z_dim}
    \vspace{-0.5em}
\end{figure}

\begin{figure*}[t]
    \centering
    \subfloat[Original SDFG with independent maps]{\includegraphics[height=1.05in]{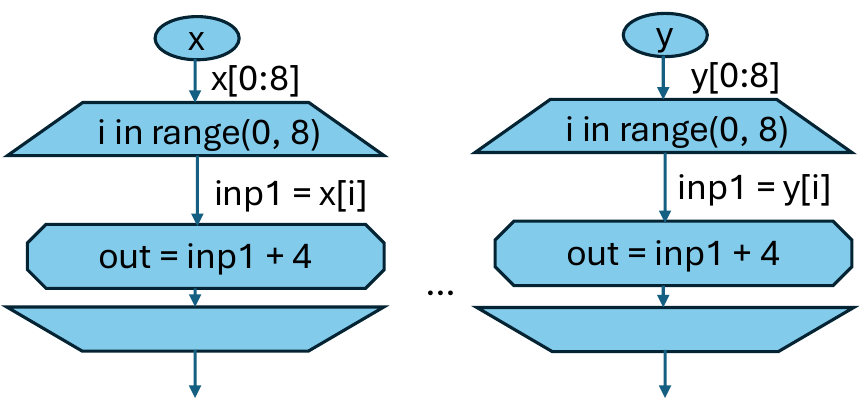}\label{fig:sdfg_dsd_pattern}}
    \hfill
    \subfloat[After horizontal map merging]{\includegraphics[height=1.05in]{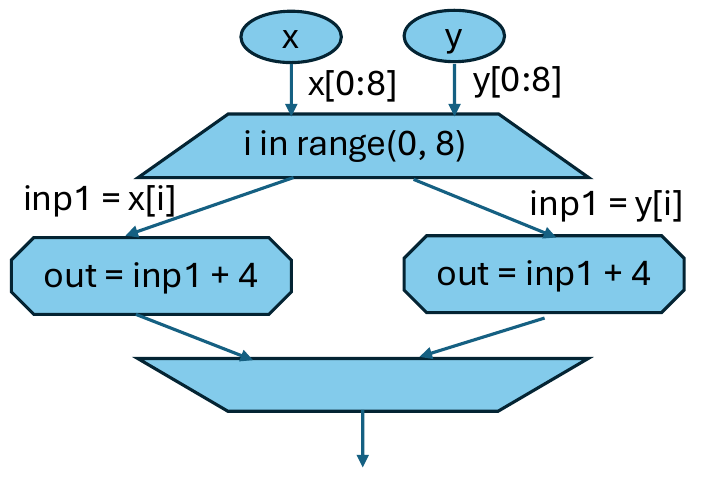}\label{fig:sdfg_stages:horizontal}}
    \hfill
    \subfloat[After array merging]{\includegraphics[height=1.05in]{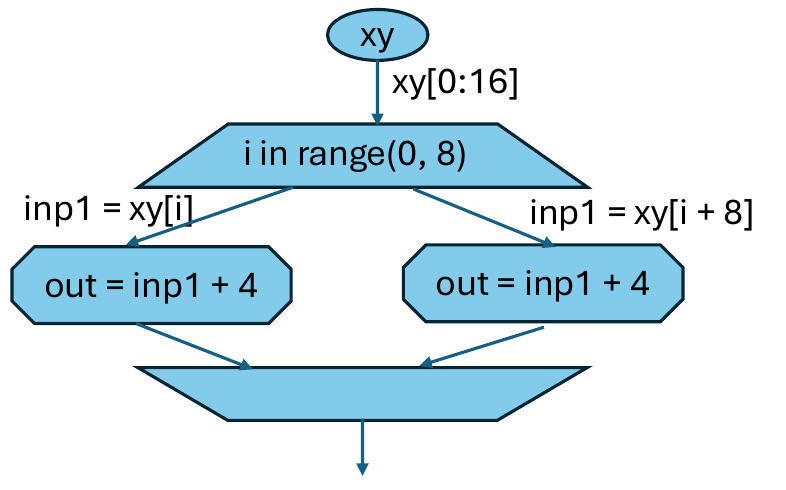}\label{fig:sdfg_stages:merging}}
    \hfill
    \subfloat[After tasklet stacking]{\hspace{0.375em}\includegraphics[height=1.05in]{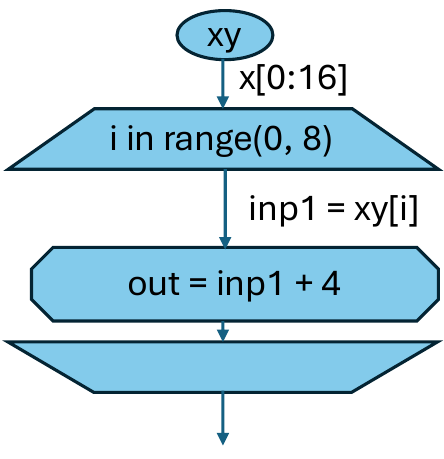}\hspace{0.375em}\label{fig:sdfg_stages:stacking}}
    \caption{SDFG transformation stages. After the three passes, the maps are merged into a single map with a single tasklet, enabling a single DSD operation.}
    \label{fig:sdfg_stages}
    \vspace{-1em}
\end{figure*}

For the Z dimension, the implementation is more intricate. A PE needs to send data to a PE that is $t_x$ hops away. Unlike networked clusters, the CS-2 does not support arbitrary peer-to-peer communication nor a message passing interface. This means that we have to manually consider that $t_x$ PEs share the same route, each of which sends to a different PE. However, the challenge is that messages must travel to their designated receivers and multiple messages must never arrive at the same router, using the same color, within a single cycle.

If we consider a group of $t_x$ PEs sending \texttt{EAST}, we configure each sender such that it receives from the \texttt{RAMP} and sends \texttt{EAST}. After receiving a control wavelet, it will start receiving from the \texttt{EAST} direction. This means that a PE will send a message, then it will send a control wavelet such that messages from the \texttt{WEST} can advance. 
Each receiver, in turn, maintains a filter that counts the number of control wavelets received. It is configured such that after a PE receives a certain number of control wavelets, it will start accepting wavelets, i.e., messages from the designated PE. We know that a message ends with a terminal control wavelet, meaning after it receives a control wavelet, it will start filtering out wavelets and another PE becomes a receiver, see Figure~\ref{fig:z_dim} for a depicted scheme.

Other potential approaches include setting up filters based on the number of total received wavelets. While this reduces the number of messages in flight (i.e., no control wavelets), we would need to reconfigure filters at runtime based on the size of the field we are communicating. Our chosen approach is more general, carries negligible cost due to pipelining, and can easily be extended to support other applications.

\subsubsection*{Combined Field Exchange}
For exchanging element properties, XY communication is followed by Z communication. Nodal properties require an additional step, since nodes on edges and corners are shared with diagonal neighbors, not just face neighbors. A generic approach would be to run a second round of XY communication after Z communication to propagate corner contributions. In LULESH, however, we accumulate results as they arrive, which allows us to instead execute Z communication first and then perform a single XY communication pass on the already-accumulated nodal values, avoiding the extra round.

\subsubsection*{Collective Communication}
Collective operations necessary for mesh-based simulations, such as allreduce, can already reuse existing libraries, such as the Cerebras SDK~\cite{cerebras_sdk} or the implementation provided by Luczynski et al.~\cite{cerebras_reduce}.

\section{Automatic Code Generation}\label{sec:compiler}
Manually writing optimized LULESH in CSL would be tedious, hence we discuss an automated approach. We parse PyLULESH~\cite{pylulesh} with DaCe~\cite{dace} v1.0.2 into an SDFG. We then divide the code into two parts: Lagrange Nodal and Lagrange Element. We use our communication library for exchanges and automatically map the sequential parts.

\subsection{SDFG Mapping}
The SDFG IR is highly amenable to conversion to performant CSL. Consider an example SDFG in Fig.~\ref{fig:sdfg_dsd_pattern}. The figure contains a SDFG map scope, where the operations inside are replicated parametric number of times. The map scope contains a tasklet connected by memlets.
This is however, \textit{exactly what a DSD intrinsic is}: the SDFG map tells us that we have an operation that will be replicated. If the map contains a single tasklet, we can match the computational pattern of a corresponding DSD intrinsic. Finally, each of the memlets, which are defined symbolically, defines a DSD. This is what makes the SDFG IR perfectly suited to convert to CSL.

Prior to code generation, we run a set of built-in DaCe optimization passes, and make others more aggressive. In particular, we modify the \texttt{TransientReuse} pass, which reuses no-longer-used memory segments of the same size, to be less conservative and work across the whole state machine.

To generate the entire program, we traverse the SDFG in topological order. If we encounter a map with a single tasklet inside, we try to lower it to a DSD intrinsic. Otherwise, we fall back to a for loop. A unique CSL optimization we define is \texttt{DSDReuse}, where memlets already declared as DSDs are reused throughout the code, saving DSD definition memory.  

\subsection{Dataflow Transformations to Optimize DSD Use}
When generating code from the original SDFG, by default only a few short (length 8) DSD operations are generated. Due to the short length, only a negligible speedup is observed. We propose an approach to modify the graph so as to make better use of DSD operations. 

It consists of three steps, visualized in Fig.~\ref{fig:sdfg_stages}---for a set of independent maps:
(1) try to merge the maps;
(2) try to merge the used arrays; and
(3) try to stack tasklets.
If all three are possible, we merge the maps into a single map with one tasklet, allowing us to express it as a singular DSD operation. In the following, we describe each component of the pass.

\subsubsection*{Horizontal Map Merging (Fig.~\ref{fig:sdfg_stages:horizontal})}
We merge independent maps, which contain tasklets that run the same code regardless of input/output data. The pass already exists in DaCe as \texttt{SubgraphFusion}, but we modify the heuristic match-and-apply behavior to only match maps with equivalent tasklets.

\subsubsection*{Array Merging (Fig.~\ref{fig:sdfg_stages:merging})}
Since the tasklets may operate on different arrays, we try to merge their address space to be consecutive. As a result, every tasklet within the map accesses the same data at different offsets.

\subsubsection*{Tasklet Stacking (Fig.~\ref{fig:sdfg_stages:stacking})}
Now that the tasklets are accessing the same data, all these tasklets can potentially be represented by a single map with a single tasklet. If possible, we replace the tasklets with a nested map, merging the symbolic expressions in the memlet. This results in a multi-dimensional map containing a single tasklet, which is equivalent to a DSD operation.

In the case of LULESH, the pass mostly merges fields with $x, y, z$ coordinates into a single composite. 
It helps reduce the number of DSDs and thus the number of instructions.

\section{Performance Evaluation}\label{sec:perf_eval}
We now turn to benchmark the full application to see how it compares against GPUs and what we would expect. All generated codes were compiled with the Cerebras CSL Compiler~\cite{cerebras_sdk} v1.4.0. Results are measured by taking the maximal PE cycles over 730{,}000 PEs. As we observe low variability in the number of cycles, we only measure multi-iteration runs and compute the mean iteration time. We compare the results with an NVIDIA A100 40 GB GPU running CUDA 13.1.

\subsection{3D Mesh Communication Library}
We benchmark the performance of our communication library to see how it compares against our model. We measure the XY and Z domain element exchange using our chosen configuration of $e_x = e_y = e_z = 1$ and increasing problem size $s$.

\begin{figure}[t]
    \centering
    \includegraphics[width=\columnwidth]{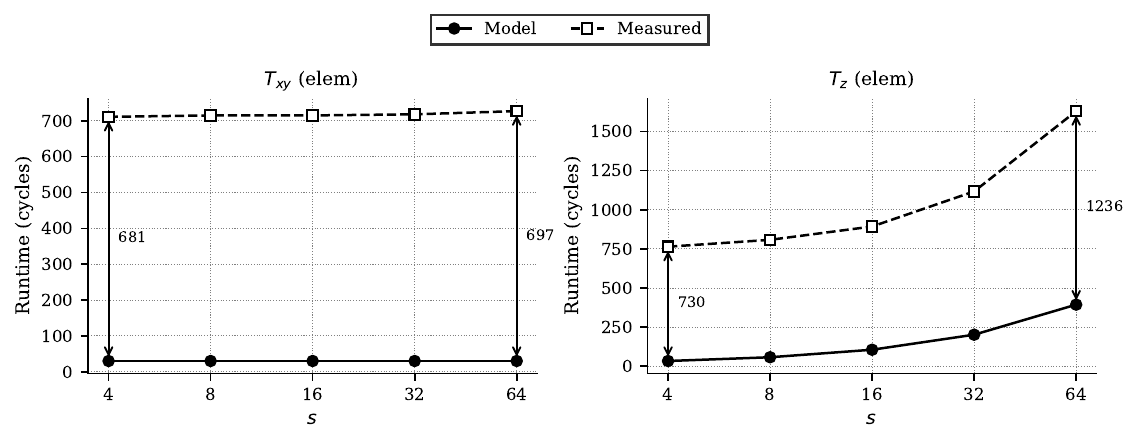}\vspace{-1em}
    \caption{Predicted runtime (model) vs. measured for XY element and Z element as a function of domain size $s$.}
    \label{fig:comm_combined}
\end{figure}

Figure~\ref{fig:comm_combined} (left) shows that the performance of the element exchange is quite far from our prediction. This can be explained by a constant setup factor for each of the four send and receive operations. 
For Z communication (Fig.~\ref{fig:comm_combined}, right), we see that for a small domain size, there is a performance gap as well. 
However, in both cases, we can see that as we increase the problem size, the difference in performance between the model and measured stays consistent.
Although there is a performance overhead, this validates our decision to use the model, since the model accurately predicts how performance changes with program parameters.

\subsection{Data-Centric Optimizations}
\begin{table}[t]
\centering
\renewcommand{\arraystretch}{1.2}
\begin{minipage}[t]{0.48\columnwidth}
\centering
\caption{Lagrange Nodal}\vspace{-0.5em}
\begin{tabular}{lrr}
\toprule
 & \textbf{Runtime} & \textbf{Size} \\
\midrule
Predicted & 8{,}724 & 9374 \\
Measured  & 16{,}969 & 17360  \\
\bottomrule
\end{tabular}

\label{tab:lagrange_nodal}
\end{minipage}
\hspace{0.02\columnwidth}
\begin{minipage}[t]{0.48\columnwidth}
\centering
\caption{Lagrange Element}\vspace{-0.5em}
\begin{tabular}{lrr}
\toprule
 & \textbf{Runtime} & \textbf{Size} \\
\midrule
Predicted & 6{,}389 & 9126 \\
Measured  & 11{,}015 & 14264 \\
\bottomrule
\end{tabular}
\label{tab:lagrange_element}
\end{minipage}
\end{table}
In Tables~\ref{tab:lagrange_nodal} and~\ref{tab:lagrange_element}, we benchmark the performance and measure the code size of our approach for each individual function. Interestingly, we can see that the measured runtime is less than $2\times$ the predicted value. 
This is due to multiple factors, mostly CSL-compiler dependent. A deep-dive of the compiled assembly shows that one main reason is that address computation when accessing domain values takes multiple cycles due to offsetting. The $T_{Seq}$ model can be adapted to support that, at the cost of it being more context-dependent and hardware-involved.

Overall, this result proves that an automated approach can accurately predict an application's runtime on a spatial dataflow architecture, as the generated code is less than 2$\times$ slower than predicted.

\subsection{End-to-end Performance}
\begin{figure}[t]
    \centering
    \includegraphics[width=.8\columnwidth,clip,trim={0cm 0.25cm 0cm 0cm}]{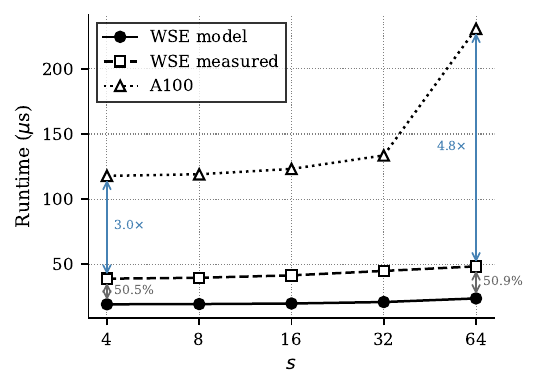}\vspace{-0.5em}
    \caption{End-to-end runtime for LULESH on CS-2 (model and measured) and A100. Arrows indicate the absolute difference between model and measured WSE runtimes, and the relative speedup of WSE over A100.}
    \label{fig:main_results}
\end{figure}

Lastly, we measure the end-to-end performance of our application and compare it with both our model and the A100 GPU. Both codes run in 32-bit precision, as the CS-2 does not support 64-bit floating point datatypes natively. The results are plotted in Fig.~\ref{fig:main_results}.

We can see that we achieve a speedup of up to 4.8$\times$ over A100. As opposed to the A100, the runtime on the WSE shows great parallel efficiency, scaling slowly due to the communication cost. If it were possible to reduce instruction memory size and run the program without subprogram PEs, an even greater speedup would be achievable.

Additionally, we can see that in a whole-program context, our model adequately represents the runtime. The accuracy of the model is consistent across domain sizes, and ranges between $50.5$ and $50.9\%$.
\section{Conclusion}

Mapping scientific applications onto spatial dataflow architectures remains challenging. Although prior work has largely focused on individual kernels, extending these efforts to full applications introduces challenging scaling constraints. In this paper, we presented a principled methodology for mapping applications to the Cerebras WSE through performance and communication modeling. We applied this approach to LULESH, an unstructured Lagrangian mesh hydrodynamics application, and reproduced the scaling behavior observed on hardware with only a constant overhead. We further showed that these models can be derived analytically from NumPy code using the SDFG representation. The same representation also enables memory optimizations such as array merging, computational stacking, and memory reuse, as well as efficient CSL code generation with DSD intrinsic operations.

Taken together, the modeling framework and code generator enable domain decomposition and computational decisions to be made systematically, yielding nontrivial speedups over same-generation GPUs despite highly disaggregated memory and long-distance communication. More broadly, by codifying the mapping process, automating much of the implementation effort, and providing reusable libraries, this work establishes a practical foundation for bringing full scientific applications to the WSE and other spatial dataflow architectures.

\section*{Acknowledgments}
Work by Lawrence Livermore National Laboratory was performed under the auspices of the U.S. Department of Energy under contract DE-AC52-07NA27344 (LLNL-CONF-2021720).
Generative AI was used to improve the wording of several sentences in this paper~\cite{sonnet46,gpt54}.

\bibliographystyle{IEEEtran}
\bibliography{citations}

\end{document}